# Current-driven domain wall motion in magnetic wires with asymmetric notches

A. Himeno, S. Kasai, and T. Ono

*Institute for Chemical Research, Kyoto University, Uji, Kyoto 611-0011, Japan*

**Abstract**

Current-driven domain wall (DW) motion in magnetic wires with asymmetric notches was investigated by means of magnetic force microscopy. It was found that the critical current density necessary for the current-driven DW motion depended on the propagation direction of the DW. The DW moved more easily in the direction along which the slope of the asymmetric notch was less inclined.

**Corresponding author:** Atsushi Himeno

Laboratory of Magnetic Materials, Division of Materials Chemistry,

Institute for Chemical Research, Kyoto University,

Gokasho, Uji, Kyoto 611-0011, Japan

*e*-mail: himeno@ssc1.kuicr.kyoto-u.ac.jp

Tel: +81-774-38-3105

Fax: +81-774-38-3109

Magnetization reversal process of microscopic ferromagnetic materials with a single magnetic domain has been researched worldwide and actively because of improving the lithographically fabrication technology. The magnetic domain configuration and the magnetization reversal process are controllable by modifying the shape of a sample. Especially, in a magnetic wire with submicron width, due to the magnetic shape anisotropy, the magnetization direction is restricted in parallel to the wire axis, and thus the state of a single magnetic domain is stabilized. The magnetization reversal of a submicron magnetic wire takes place by the propagation of a magnetic domain wall (DW). Several experimental attempts have been reported to control the nucleation and the propagation of a magnetic DW in submicron magnetic wires.[1-7] In a previous report,[6] we have shown that the depinning field of the DW from the asymmetric notch artificially introduced into a submicron magnetic wire depends on the propagation direction of the DW, and that the asymmetric notch works as an asymmetric potential barrier against the propagation of a magnetic DW. This phenomenon should be called a "magnetic ratchet effect". A similar effect was found in the experiment on submicron magnetic wires with a triangular structure using a magneto-optical Kerr effect.[7]

In this letter, we present a study on the current-driven DW motion[8-17] in submicron magnetic wires with asymmetric notches. It was confirmed that the magnetic ratchet effect can be generated not only by a magnetic field but also by an electric current through the wire.

Samples were fabricated by lift-off method in combination with an electron beam lithography onto thermally oxidized Si substrates. Figure 1 shows a schematic illustration of a top view of the sample. The sample consists of an arched $Ni_{81}Fe_{19}$ (Py) wire with asymmetric notches (25 nm in thickness and 200 μm in radius) and four current-voltage electrodes made of Au. It has been shown that, for the magnetic wire of similar dimension to the present sample, the depinning fields for the rightward propagation and for the leftward propagation were about 100 Oe and 50 Oe, respectively.[6] Because of the arched shape of the Py wire, a magnetic DW can be introduced into a center of the Py wire by applying an external magnetic field ($H_{ext}$) in one direction.[17]

At first, $H_{ext}$ of 3 kOe was applied in order to saturate the magnetization in one direction, and then $H_{ext}$ was decreased to zero. After that, the atomic force microscopy (AFM) and the magnetic force microscopy (MFM) observations were carried out in the absence of $H_{ext}$. The existence of a single DW at the vicinity of the center of the arched wire was observed as shown in Fig. 2(a) and (b). The DW is imaged as a dark contrast, which corresponds to the stray field from a negative magnetic charge, indicating that a tail-to-tail DW is realized. Though the following experiment result was on the tail-to-tail DW, the same result was obtained also for the head-to-head DW. After the observation of Fig. 2(b), a pulsed current with the duration of 2 μs was applied through the Py wire from left to right. The MFM observation was carried out after applying each pulse with increasing the intensity of the pulsed current. It was confirmed that the critical current density for the current-driven DW motion was $7.2 \times 10^{11}$ A/m$^2$ for the leftward propagation of the DW.[18] This is comparable to the critical current density for the simple magnetic wire of similar dimension.[12] Figures 2(c)-(e) are successive MFM images with one pulsed current of $7.2 \times 10^{11}$ A/m$^2$ between each consecutive image. Each pulsed current displaced the DW notch by notch to the leftward direction. It should be noted that the DW of the vortex type was pinned at the widest part of the asymmetric notch.

We tried to determine the critical current density for the rightward propagation by switching the current polarity. However, the DW did not move rightward up to the current density of $8.9 \times 10^{11}$ A/m$^2$. Above this current density, the multi-domain structure appeared as shown Fig. 2(f). This is the indication that the sample temperature exceeded the Curie temperature of Py by the Joule heating because of the high current density.[13]

Therefore, we found a clear difference of the critical current density between the leftward propagation and the rightward propagation. The DW moved more easily in the direction along which the slope of the asymmetric notch was less inclined. This situation is the same in the case of the DW motion in magnetic wires with asymmetric notches by an external magnetic field.[6] Thus, it is concluded that the asymmetric notch works as an asymmetric potential not only for the field-driven DW motion but also for the current-driven DW motion.

In conclusion, we have demonstrated the magnetic ratchet effect for the current-driven DW motion in submicron magnetic wires with asymmetric notches. The current density to depin the DW from the asymmetric notch depends on the DW propagation direction. The present result certifies that the DW motion can be controlled by the artificial structure even for the current-driven DW motion.

## Acknowledgement

The present work was partly supported by MEXT Grants-in-Aid for Exploratory Research, MEXT Grants-in-Aid for Scientific Research in Priority Areas,

JSPS Grants-in-Aid for Scientific Research, and Industrial Technology Research Grant Program in '05 from NEDO of Japan.

**Figure captions**

**Figure 1.** Schematic illustration of a top view of a sample consists of an arched $Ni_{81}Fe_{19}$ (Py) wire and Au electrodes. A main body of the arched Py wire has 10 asymmetric notches.

**Figure 2.** **(a)** AFM image of the arched Py wire with asymmetric notches. **(b)** MFM image of the initial magnetization configuration. A tail-to tail DW was observed, which was imaged as a dark contrast. **(c)-(e)** Successive MFM images with one pulsed current applied between each consecutive image. The pulsed current with the duration of 2 μs flowed through the Py wire from left to right. The applied current density was $7.2 \times 10^{11}$ A/m$^2$. Each pulsed current displaced the DW notch by notch. **(f)** MFM images after the application of a pulsed current with the current density of $8.9 \times 10^{11}$ A/m$^2$ in the opposite direction. Multi-domain structure appeared in the Py wire.

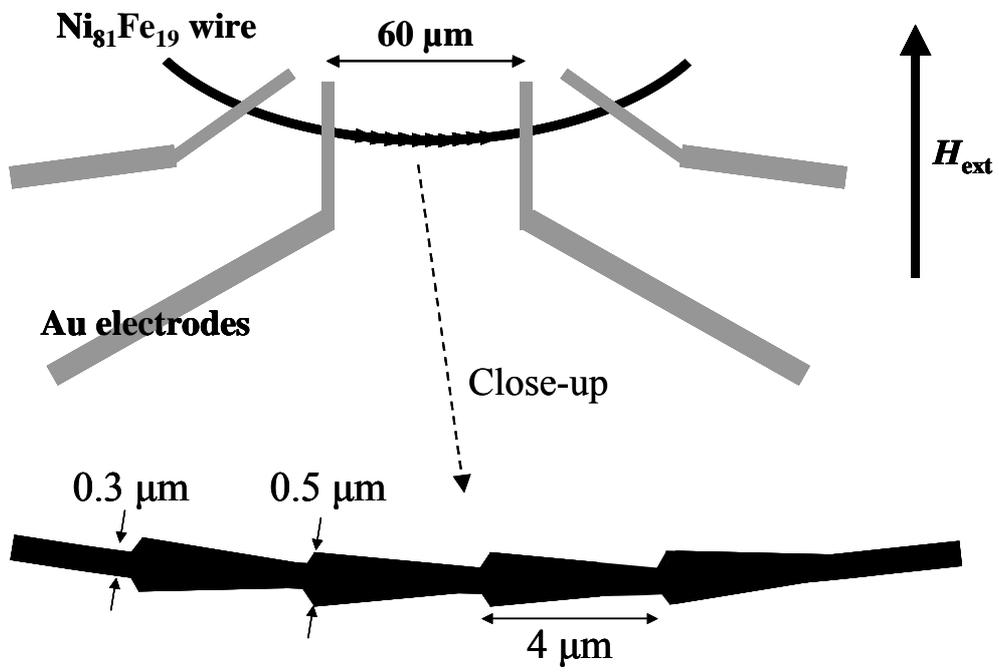

A. Himeno *et al*. - Figure 1

(a) AFM image

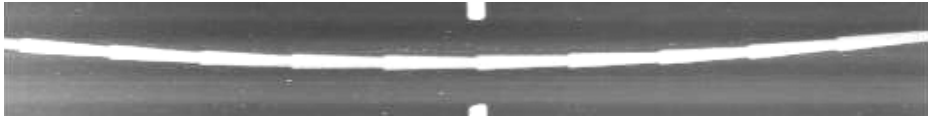

(b) Initial configuration

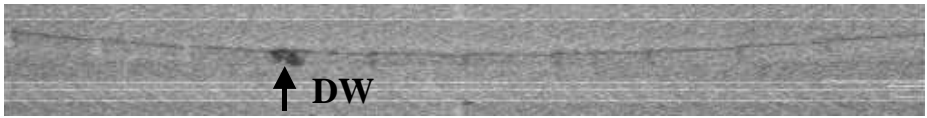

↑ DW

(c) 7.2 × 10$^{11}$ A/m$^2$     **current** →

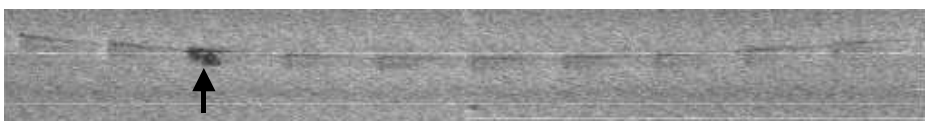

↑

(d) 7.2 × 10$^{11}$ A/m$^2$     →

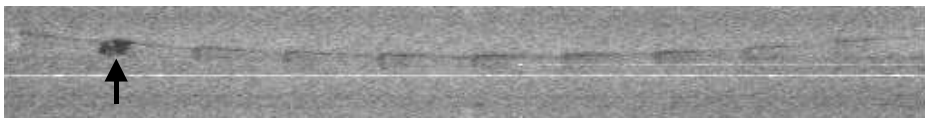

↑

(e) 7.2 × 10$^{11}$ A/m$^2$     →

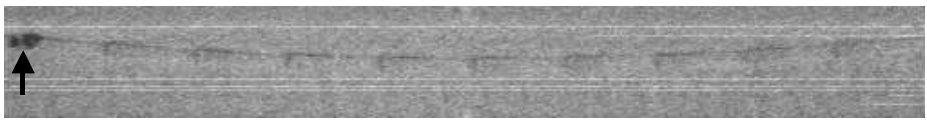

↑

(f) 8.9 × 10$^{11}$ A/m$^2$     **current** ←

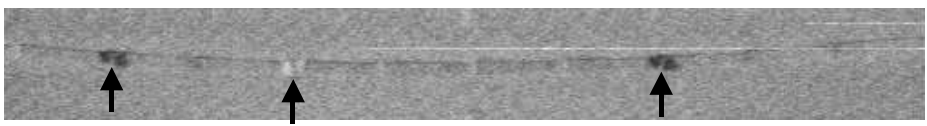

↑   ↑        ↑

5 μm

A. Himeno *et al*. - Figure 2